\documentclass[aps,twocolumn,pra,superscriptaddress,tightenlines]{revtex4-1}
\usepackage{CJK}
\usepackage{amsfonts}
\usepackage{graphicx}
\usepackage{hyperref}
\usepackage{amsmath}
\usepackage{amssymb}
\usepackage[usenames, dvipsnames]{color}
\usepackage{subfigure}
\usepackage{lipsum}

\begin{document}

\title{Spin relaxation in inhomogeneous magnetic fields with depolarizing
boundaries}
\author{Yue Chang}
\email{yuechang7@gmail.com}
\author{Shuangai Wan}
\author{Shichao Dong}
\author{Jie Qin}
\email{j.qin@aliyun.com}

\affiliation{Beijing Automation Control Equipment Institute, Beijing 100074,
China}
\affiliation{Quantum Technology R$\&$D Center of China Aerospace
Science and Industry Corporation, Beijing 100074, China}

\begin{abstract}
Field-inhomogeneity-induced relaxation of atomic spins confined in vapor
cells with depolarizing walls is studied. In contrast to nuclear spins, such
as noble-gas spins, which experience minimal polarization loss at cell
walls, atomic spins in uncoated cells undergo randomization at the
boundaries. This distinct boundary condition results in a varied dependence
of the relaxation rate on the field gradient. By solving the Bloch-Torrey
equation under fully depolarizing boundary conditions, we illustrate that
the relaxation rate induced by field inhomogeneity is more pronounced for
spins with a smaller original relaxation rate (in the absence of the
inhomogeneous field). We establish an upper limit for the relaxation rate
through calculations in the perturbation regime. Moreover, we connect it to
the spin-exchange-relaxation-free magnetometers, demonstrating that its
linewidth is most sensitive to inhomogeneous fields along the magnetometer's
sensitive axis. Our theoretical result agrees with the experimental data for
cells subjected to small pump power. A deviation in high input-power
scenarios arises from pump field attenuation, resulting in a non-uniformly
distributed light shift that behaves like an inhomogeneous magnetic field.
\end{abstract}

\maketitle

\section{Introduction}

Atomic sensors such as optically-pumped magnetometers \cite%
{Budker2003,Kominis2003,Budker2007} and devices based on nuclear magnetic
resonance \cite%
{Walker2016,Donley2010,Walker2015,PhysRevLett.94.123001,Kornack2005},
leverage the response of electron or nuclear spins to external magnetic
fields for precision measurement. In the presence of an inhomogeneous
magnetic field, these spins undergo dephasing precession, leading to a
reduction in their longitudinal and transverse relaxation times. For spins
in the gas phase, the effect of the inhomogeneous field on relaxation
depends on the boundary conditions. Under non-depolarizing boundary
conditions, such as noble-gas spins confined in a cell, there has been
extensive theoretical and experimental study on the influence of
inhomogeneous static and oscillating fields \cite%
{PhysRevA.37.2877,PhysRevA.38.5092,PhysRevA.41.3672,PhysRevA.41.2631,PhysRevA.84.053411,PhysRevA.74.063406,PhysRevA.104.042819,PhysRevLett.113.163002}%
, particularly in the diffusion regime where the system size exceeds the
particles' mean free path. Under the perturbation condition where the
inhomogeneity is much smaller than the gas diffusion effect characterized by
the diffusion constant $D_{n}$ and system length $L$, analytical expressions
for relaxation times have been derived. Particularly, for scenarios
involving a high longitudinal (z direction) magnetic field or large gas
pressure, the influence of the inhomogeneous field on longitudinal
relaxation can be neglected. Then, the inhomogeneity-dependent transverse
decay rate $\Gamma _{2}^{n}$ of nuclear spins in a cubic vapor cell with a
length of L can be approximated as \cite{PhysRevA.41.2631}%
\begin{equation}
\Gamma _{2}^{n}=\frac{\gamma _{n}^{2}L^{4}\left\vert \nabla B_{z}\right\vert
^{2}}{120D_{n}},  \label{1}
\end{equation}%
where $\gamma _{n}$ is the gyromagnetic ratio of the nuclear spin. Here, a
linear dependence of $B_{z}\ $on the position $\vec{r}$ is assumed, namely, $%
\partial _{x,y,z}B_{z}$ is a constant. The investigation of second-order
field gradients has also been addressed in a prior study \cite%
{PhysRevLett.113.163002}. We note that throughout existing literature,
nondepolarizing boundary conditions have been consistently employed, as
wall-collision-induced spin randomization, particularly for nuclei like
commonly used noble gases, is typically considered negligible in most
scenarios \cite{Wu2021}. However, when dealing with atomic spins such as
alkali-metal atomic spins, which experience near-total polarization loss at
uncoated cell walls \cite{Wu2021}, the assumptions and results established
for nondepolarizing-boundary spins needs to be reexamined.

In this paper, we explore the relaxation dynamics of atomic spins in the
presence of an inhomogeneous magnetic field characterized by a linear
gradient. A fully depolarizing boundary condition \cite%
{Wu2021,PhysRevA.109.023113} is adopted for atoms diffusing in a cubic vapor
cell. By solving the Bloch-Torrey equation \cite%
{PhysRevLett.113.163002,PhysRevA.109.023113} of atomic spins in the
perturbation regime, we demonstrate that different from expectations based
on nuclear spin behavior, the inhonogeneity-induced transverse relaxation
rate $\Gamma _{2}^{a}$ does not surpass%
\begin{equation}
\Gamma _{2}^{a}=\frac{\gamma _{a}^{2}\left\vert \nabla B_{z}\right\vert
^{2}L^{4}}{D_{a}}\frac{15-\pi ^{2}}{48\pi ^{4}},
\end{equation}%
where $D_{a}$ and $\gamma _{a}$ are the diffusion constant and gyromagnetic
ratio for the atomic spins. Therefore, substituting $\gamma _{n}$ and $D_{n}$
with $\gamma _{a}$ and $D_{a}$ in Eq. (\ref{1}) would yield an estimation of
the relaxation rate that is at least eight times larger than the correct
value. This arises from the fine spatial features of spin polarization.
Unlike the uniform distribution of spin polarizations observed with
non-depolarizing boundaries (in the absence of field inhomogeneity and pump
light attenuation), spin polarizations under fully depolarizing boundary
conditions are concentrated in a smaller volume away from the cell walls.
This finer spatial confinement, where atoms sample a smaller region, leads
to reduced sensitivity to magnetic field inhomogeneity. Furthermore, our
study demonstrates that the inhomogeneity-induced atomic relaxation is not
solely dependent on the field gradient $\gamma _{{a}}\left\vert \nabla B_{{z}%
}\right\vert $, system size $L$, and the diffusion constant $D_{a}$. It is
also dependent on the original longitudinal or transverse relaxation rate $%
\Gamma _{0}^{a}$ when the field inhomogeneity is absent. Across a wide range
of field gradients, a smaller $\Gamma _{0}^{a}$ corresponds to a more
significant increase in $\Gamma _{2}^{a}$ due to the diminished
effectiveness of motional narrowing \cite%
{doi:10.1143/JPSJ.9.316,PhysRevLett.96.117203}.

We also establish a practical link to our study by investigating the
linewidth broadening of spin-exchange-relaxation-free (SERF) magnetometers
\cite{Happer1977,Allred2002,PhysRevA.71.023405}, whose linewidth is narrow
and thus is susceptible to the field inhomogeneity. In absence of a large
longitudinal magnetic field, both longitudinal and transverse relaxation can
experience significant alterations due to the inhomogeneous field, resulting
in the simultaneous broadening of a SERF magnetometer's linewidth. Beyond
the dependence on $\Gamma _{0}^{a}$, linewidth broadening responds
differentially to magnetic fields along three directions. The impact from
the sensitive-axis direction (designated as the $y$ direction, aligned with
the magnetic field that requires measurement) is the most pronounced, while
inhomogeneous fields in the other two directions (the longitudinal direction
and the signal direction denoted as the $x$ direction) contribute equally to
the linewidth. Our theoretical results align with experimental data,
exhibiting agreement at lower optical pumping powers. Deviations observed at
higher input powers suggest the need to consider the attenuation of pump
light \cite{PhysRevA.58.1412,PhysRevA.109.023113} and the corresponding
non-uniformly distributed light shift \cite%
{PhysRevA.58.1412,PhysRevA.99.063411}.

This paper is structured as follows: following the introduction, Section II
undertakes an analysis of transverse relaxation induced by the inhomogeneous
longitudinal magnetic field. This is accomplished through the solution of
Bloch-Torrey equations under the depolarizing-boundary condition,
elucidating the dependence of the transverse decay rate on the linear field
gradient within the perturbation regime. Subsequently, Section III
establishes a connection between this analysis and the linewidth broadening
in SERF magnetometers. Inhomogeneous magnetic fields in three directions are
considerd, and our theoretical results agrees with experimental data when
the non-uniformly distributed light shift can be neglected. The paper
concludes with Section IV, presenting a summary and outlooks for future
research.

\section{Transverse relaxation due to inhomogeneous magnetic field $%
B_{z}\left( \vec{r}\right) $}

In this section, we delve into the analysis of the transverse relaxation of
atomic spins under the influence of an inhomogeneous magnetic field in the $%
z $ direction. Without loss of generality, we consider a cubic vapor cell,
where the atomic spins are fully depolarized at the walls. This is
applicable to atoms such as alkali-metal atoms in vapor cells lacking
antirelaxation coatings \cite{Wu2021}. Denoting the polarization of atomic
spins along three directions as $P_{x,y,z}(\vec{r},t)$, the Bloch-Torrey
equation of $P_{+}(\vec{r},t)=P_{x}(\vec{r},t)+iP_{y}(\vec{r},t)$ under a
longitudinal magnetic field $B_{z}\left( \vec{r}\right) $ is \cite%
{PhysRevLett.113.163002,PhysRevA.109.023113}%
\begin{equation}
\partial _{t}P_{+}(\vec{r},t)=D_{a}\nabla ^{2}P_{+}(\vec{r},t)-i\gamma
_{a}B_{z}\left( \vec{r}\right) P_{+}(\vec{r},t)-\Gamma _{0}^{a}P_{+}(\vec{r}%
,t),  \label{7}
\end{equation}%
where $\Gamma _{0}^{a}$ is the transverse decay rate without field
inhomogeneities and at the cell walls $\left. P_{+}(\vec{r},t)\right\vert
_{S}=0$. Note that the validity of the diffusion equation holds when the
system size $L$ exceeds the mean-free path of the particle \cite%
{landau2013fluid}. In commonly used alkali-metal-atom vapor cells, where the
average atomic velocity is larger than $10^{2}$m/s, the diffusion condition
is readily satisfied as long as $3D/L<$ $10^{2}$m/s---a condition
consistently met throughout this paper.

The transverse relaxation rate can be determined through free induction
decay \cite{slichter2013principles}, where a $\pi /2$ pulse along the $x$ or
$y$ direction is applied to flip the longitudinal polarization $P_{z}(\vec{r}%
,t)$ to the $y$ or $x$ direction. Assuming it is a $y$-direcion pulse, the
initial conditions are $P_{y}(\vec{r},0)=0$ while $P_{x}(\vec{r},0)$ is
determined by the longitudinal distribution $P_{z}(\vec{r},t)$ at the
long-term limit:
\begin{equation}
D_{a}\nabla ^{2}P_{x}(\vec{r},0)-\Gamma _{0}^{a}P_{x}(\vec{r},0)+R=0,
\label{2}
\end{equation}%
where $R$ is polarization rate, for instance, the optical-pumping rate \cite%
{RevModPhys.44.169,PhysRevA.58.1412}, and $P_{x}(\vec{r},0)=0$ at the cell
walls. Here, to provide clear physics insights, we have minimized the number
of parameters by assuming equivalence between the longitudinal decay rate
and the transverse one, denoted as $\Gamma _{0}^{a}$. Nevertheless, it is
important to acknowledge that the analysis can be extended to other
scenarios and similar conclusions can be obtained. Subsequently, we first
focus on the perturbation regime characterized by small inhomogeneities.

\subsection{Perturbation regime}

Under the fully depolarizing boundary condition, for cubic cells the
polarization $P_{+}(\vec{r},t)$ can be expanded as%
\begin{equation}
P_{+}(\vec{r},t)=\left( \frac{L}{2}\right)
^{3/2}\sum\limits_{mnl=1}C_{mnl}\left( t\right) \psi _{mnl}^{\left( 0\right)
}\left( x,y,z\right) ,
\end{equation}%
where $\psi _{mnl}^{\left( 0\right) }\left( x,y,z\right) =\phi _{m}\left(
x\right) \phi _{n}\left( y\right) \phi _{l}\left( z\right) $ with%
\begin{equation}
\phi _{m}\left( x\right) =\sqrt{\frac{2}{L}}\sin \left( \frac{m\pi }{L}%
x\right) ,m=1,2,..,
\end{equation}%
is the eigenmodes of the diffusion equation in absence of the field
inhomogeneity and the initial condition $C_{mnl}\left( 0\right) $ is
determined by Eq. (\ref{2}) as%
\begin{equation}
C_{mnl}\left( 0\right) =\frac{64}{mnl\pi ^{3}}\frac{R}{\Gamma _{mnl}^{\left(
0\right) }}.
\end{equation}%
Here,%
\begin{equation}
\Gamma _{mnl}^{\left( 0\right) }=\Gamma _{0}^{a}+D_{a}\left( \frac{\pi }{L}%
\right) ^{2}\left( m^{2}+n^{2}+l^{2}\right)
\end{equation}%
is the decay rate for the eigenmodes $\phi _{m}\left( x\right) \phi
_{n}\left( y\right) \phi _{l}\left( z\right) $ in absence of the
inhomogeneous field.

The presence of the inhomogeneous field induces position-dependent
precession frequencies, giving rise to additional decay channels attributed
to the dephasing of transverse spin polarization. In the perturbation regime
where%
\begin{equation}
\left\vert \frac{\gamma _{a}\int dV\text{ }\psi _{m^{\prime }n^{\prime
}l^{\prime }}^{\left( 0\right) }\left( x,y,z\right) B_{z}\left( \vec{r}%
\right) \psi _{mnl}^{\left( 0\right) }\left( x,y,z\right) }{\Gamma
_{mnl}^{\left( 0\right) }-\Gamma _{m^{\prime }n^{\prime }l^{\prime
}}^{\left( 0\right) }}\right\vert \ll 1,  \label{3}
\end{equation}%
to the second order of the inhomogeneity, the coefficient $C_{mnl}\left(
t\right) $ can be written in the form $C_{mnl}\left( t\right) =C_{mnl}\left(
0\right) e^{-i\omega _{mnl}t-\Gamma _{mnl}t}$, where%
\begin{equation}
\omega _{mnl}^{\left( 1\right) }=\gamma _{a}\int dV\text{ }\psi
_{mnl}^{\left( 0\right) }\left( x,y,z\right) B_{z}\left( \vec{r}\right) \psi
_{mnl}^{\left( 0\right) }\left( x,y,z\right)
\end{equation}%
is an effective precession frequency and $\Gamma _{mnl}=\Gamma
_{mnl}^{\left( 0\right) }+\Gamma _{mnl}^{\left( 2\right) }$ is the
mode-dependent transverse relaxation rate with the inhomogeneity-induced
decay rate as%
\begin{equation}
\Gamma _{mnl}^{\left( 2\right) }=\gamma _{a}^{2}\sum\limits_{m^{\prime
}n^{\prime }l^{\prime }}\frac{\left\vert \int dV\text{ }\psi _{m^{\prime
}n^{\prime }l^{\prime }}^{\left( 0\right) }\left( x,y,z\right) B_{z}\left(
\vec{r}\right) \psi _{mnl}^{\left( 0\right) }\left( x,y,z\right) \right\vert
^{2}}{\Gamma _{mnl}^{\left( 0\right) }-\Gamma _{m^{\prime }n^{\prime
}l^{\prime }}^{\left( 0\right) }}.
\end{equation}

For linear gradient $B_{z}\left( \vec{r}\right) =gx$, the pertubeation
condition (\ref{3}) becomes%
\begin{equation}
\frac{16\gamma _{a}gL^{3}}{27\pi ^{4}D_{a}}\ll 1,  \label{6}
\end{equation}%
the effective frequency $\omega _{1}\equiv \omega _{mnl}^{\left( 1\right)
}=\gamma _{a}gL/2$\textbf{\ }\textit{is independent of}\textbf{\ }$mnl$, and
the decay rate $\Gamma _{mnl}^{\left( 2\right) }$ is only dependent on $m$.
Since $C_{mnl}\left( 0\right) $ is inversely proportional to $mnl$, $P_{+}(%
\vec{r},t)$ is dominated by modes with small $m,$ $n,$ and $l$. The first
three $\Gamma _{mnl}^{\left( 2\right) }$ are%
\begin{equation}
\Gamma _{1nl}^{\left( 2\right) }=\frac{\gamma _{a}^{2}g^{2}L^{4}}{D_{a}}%
\frac{15-\pi ^{2}}{48\pi ^{4}},  \label{8}
\end{equation}%
\begin{equation}
\Gamma _{2nl}^{\left( 2\right) }=\frac{\gamma _{a}^{2}g^{2}L^{4}}{D_{a}}%
\frac{15-4\pi ^{2}}{768\pi ^{4}},
\end{equation}%
\begin{equation}
\Gamma _{3nl}^{\left( 2\right) }=-\frac{\gamma _{a}^{2}g^{2}L^{4}}{D_{a}}%
\frac{3\pi ^{2}-5}{1296\pi ^{4}}.  \label{5}
\end{equation}%
And the transverse polarization can be approximated as%
\begin{equation}
P_{+}(\vec{r},t)\approx \left( \frac{L}{2}\right) ^{\frac{3}{2}}e^{-\frac{i}{%
2}\omega _{1}t}\sum\limits_{mnl=1}^{3}C_{mnl}\left( 0\right) \psi
_{mnl}^{\left( 0\right) }\left( x,y,z\right) e^{-\Gamma _{mnl}t},
\end{equation}%
which immidiatly gives the average transverse polarization%
\begin{eqnarray}
\bar{P}_{+}\left( t\right) &\equiv &\frac{1}{L^{3}}\int dV\text{ }P_{+}(\vec{%
r},t)  \notag \\
&\approx &\frac{512}{\pi ^{6}}Re^{-i\omega _{1}t}\sum\limits_{mnl=1,3}\frac{%
e^{-\Gamma _{mnl}t}}{\left( mnl\right) ^{2}\Gamma _{mnl}^{\left( 0\right) }}.
\end{eqnarray}%
The relaxation rate is determined by the time when the output decays to $1/e$
from the initial value. Therefore, we normalize $\bar{P}_{+}\left( t\right) $
to the initial value, and acquire%
\begin{eqnarray}
\bar{P}_{+}^{Nor}\left( t\right) &\equiv &\bar{P}_{+}\left( t\right) /\bar{P}%
_{+}\left( 0\right)  \notag \\
&=&e^{-i\omega _{1}t}\sum\limits_{mnl=1,3}w_{mnl}e^{-\Gamma _{mnl}t},
\label{4}
\end{eqnarray}%
where
\begin{equation}
w_{mnl}=\left[ \left( mnl\right) ^{2}\Gamma _{mnl}^{\left( 0\right)
}\sum\limits_{mnl=1,3}\frac{1}{\left( mnl\right) ^{2}\Gamma _{mnl}^{\left(
0\right) }}\right] ^{-1}
\end{equation}%
is the weight for each mode that decreases as $m$, $n$, or $l$ increases.

\begin{figure}[tbp]
\begin{center}
\includegraphics[width=0.85\linewidth]{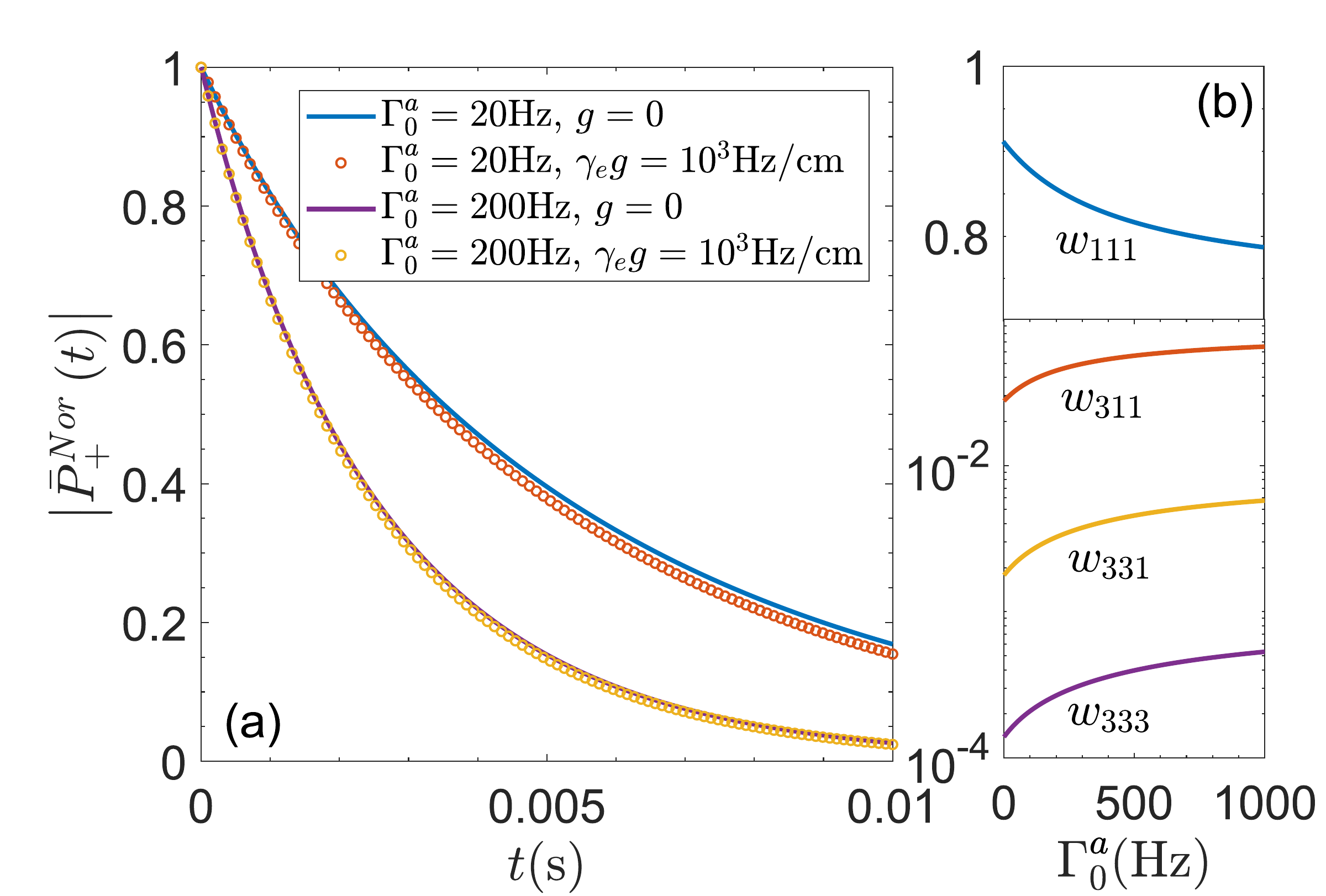}
\end{center}
\caption{(a) Normalized transvers polarization $\left\vert \bar{P}%
_{+}^{Nor}\left( t\right) \right\vert $ and (b) weight $w_{mnl}$ for
different modes. In (a), the decay rate $\Gamma _{0}^{a}=20$Hz for the two
upper curves while $\Gamma _{0}^{a}=200$Hz for the two lower curves. The
solid curves are for $\protect\gamma _{a}g=0$, while the circles represent $%
\left\vert \bar{P}_{+}^{Nor}\left( t\right) \right\vert $ with a linear
gradient of $\protect\gamma _{a}g=10^{3}$Hz/cm. With this field gradient $%
\protect\gamma _{a}g=10^{3}$Hz/cm, the perturbation condition reads $16%
\protect\gamma _{a}gL^{3}/\left( 27\protect\pi ^{4}D_{e}\right) =0.24$.}
\label{fig1}
\end{figure}

Two key insights can be extracted from Eq. (\ref{4}): (1) given that $\Gamma
_{3nl}^{\left( 2\right) }$ is negative, as indicated in Eq. (\ref{5}), the
overall inhomogeneity-induced transverse relaxation rate is smaller than $%
\Gamma _{1nl}^{\left( 2\right) }$. This relaxation rate $\Gamma
_{1nl}^{\left( 2\right) }$ is approximately one order of magnitude smaller
than the gradient dependence observed in the relaxation rate for nuclear
spins (Eq. (\ref{1})); (2) The contribution of the field inhomogeneity to
the transverse relaxation rate is also dependent on its original decay rate $%
\Gamma _{0}^{a}$ (through the weight $w_{mnl}$) when the field inhomogeneity
is absent. This relationship is illustrated in Fig. \ref{fig1}(a), where $%
\left\vert \bar{P}_{+}^{Nor}\left( t\right) \right\vert $ is plotted for
varying rate of $\Gamma _{0}^{a}$ while other parameters are fixed as $%
D_{a}=0.2$cm$^{2}$/s and $L=0.2$cm. With larger relaxation rate $\Gamma
_{0}^{a}$, the increase in the decay rate induced by the field inhomogeneity
is smaller. This is attributed to the larger atomic population in modes with
finer spatial features, as shown in Fig. \ref{fig1}(b), where the difference
in weights $w_{mnl}$ for various modes diminishes with an increase in $%
\Gamma _{0}^{a}$. In these modes, atoms can't travel as far before their
polarization decays, resulting in them sampling a smaller region and being
less sensitive to the field gradient, This leads to a smaller decay rate $%
\Gamma _{mnl}^{\left( 2\right) }$ for higher $mnl$ values. This also
explains why spins under non-depolarizing conditions are more sensitive to
the inhomogeneous field: they occupy fewer of the finer-spatial modes. We
note that it does not contradict the motional narrowing effect \cite%
{doi:10.1143/JPSJ.9.316,PhysRevLett.96.117203}: when $D_{a}$ is invariant,
the motional narrowing for each mode remains the same; when $D_{a}$
increases, fewer fine-spatial modes are populated, but the decay rate for
each mode is reduced, leading to a smaller overall relaxation rate.

\subsection{Beyond the perturbation regime}

\begin{figure}[tbp]
\begin{center}
\includegraphics[width=0.85\linewidth]{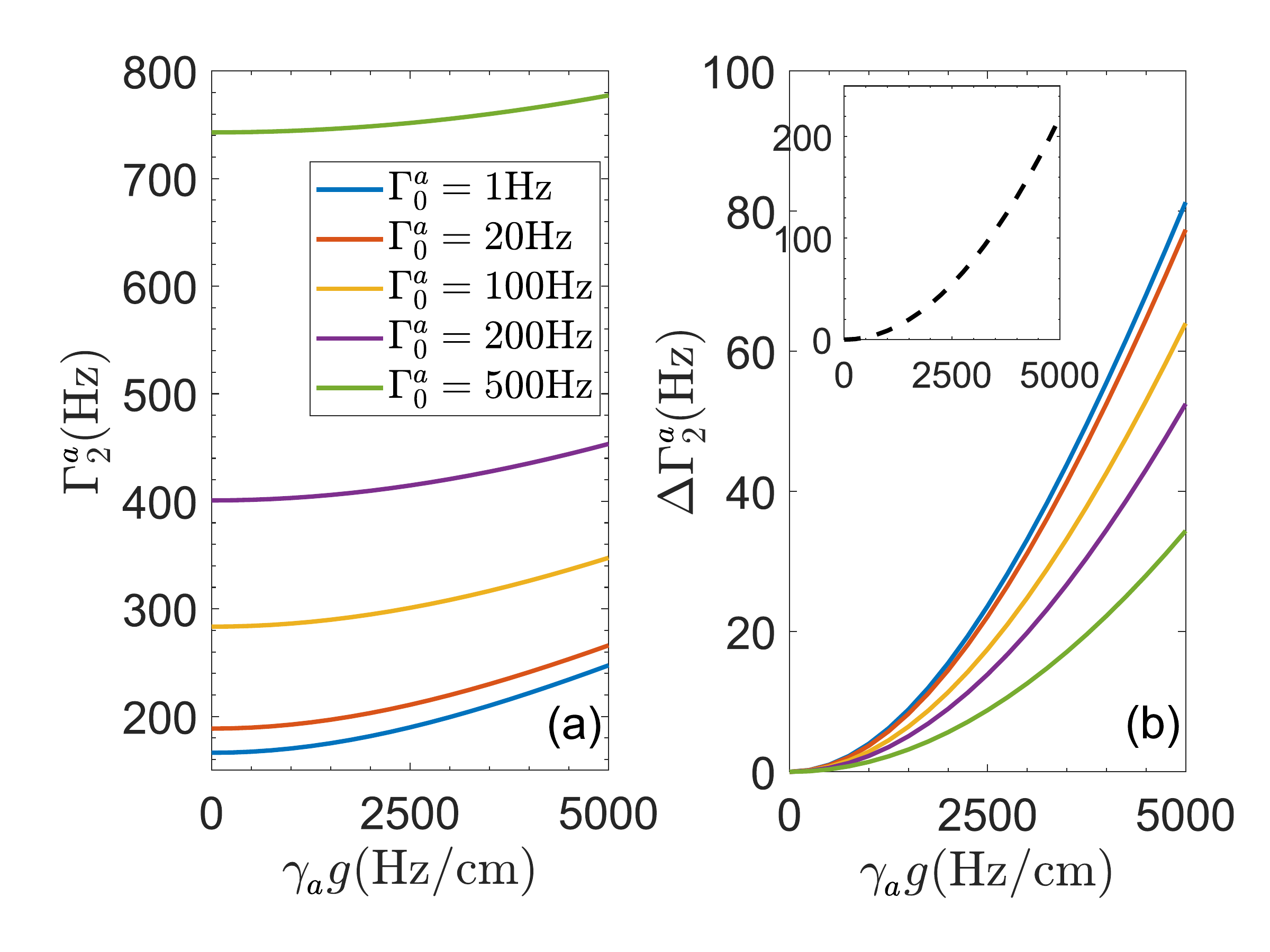}
\end{center}
\caption{(a) Transverse relaxation rate $\Gamma _{2}^{a}$ as a function of
the linear gradient $\protect\gamma _{a}g$ and (b) the contribution $\Delta
\Gamma _{2}^{a}\equiv \Gamma _{2}^{a}\left( g\right) -\Gamma _{2}^{a}\left(
g=0\right) $ from the field inhomogeneity. In both cases in (a) and (b), the
decay rate $\Gamma _{0}^{a}$ varies at values of $1$, $20$, $100$, $200$,
and $500$Hz. In (a), $\Gamma _{0}^{a}\ $increases from lower to upper
curves, while it decreases from lower to upper curves in (b), illustrating
that the influence of the inhomogeneous field becomes smaller with larger $%
\Gamma _{0}^{a}$. The inset in (b) gives a upper bound of $\Delta \Gamma
_{2}^{a}$ from the perturbation theory.}
\label{fig2}
\end{figure}

With larger field gradient, the perturbation condition (\ref{6}) is no
longer hold. In this case, we can numerically solve the Eq. (\ref{7}) under
the initial condition (\ref{2}) and the boundary condition $\left. P_{+}(%
\vec{r},t)\right\vert _{S}=0$. Denoting the overal ralaxation rate $\Gamma
_{2}^{a}$ as the inverse of the effective transverse relaxation time $T_{2}$
that is defined by%
\begin{equation}
\bar{P}_{+}^{Nor}\left( T_{2}\right) =\bar{P}_{+}^{Nor}\left( 0\right)
e^{-1},
\end{equation}%
we illustrate $\Gamma _{2}^{a}$ as a function of the gradient $\gamma _{a}g$
in Fig. \ref{fig2}(a), with $\Gamma _{0}^{a}=1$, $20$, $100$, $200$, $500$Hz
from lower to upper curves. Other parameters are $D_{a}=0.2$cm$^{2}$/s and $%
L=0.2$cm (the value of $R$ does not affect the relaxation rate since it does
not appear in the normalized polarization $\bar{P}_{+}^{Nor}$). As the
original transverse relaxation rate $\Gamma _{0}^{a}$ increases, the overal $%
\Gamma _{2}^{a}$ becomes larger, but the contribution from the field
gradient is reduced. This is illustrated in Fig. \ref{fig2}(b), where $%
\Delta \Gamma _{2}^{a}$ represents the relaxation-rate difference with and
without the field inhomogeneity, i.e., $\Delta \Gamma _{2}^{a}\equiv \Gamma
_{2}^{a}\left( g\right) -\Gamma _{2}^{a}\left( g=0\right) $. This
observation aligns with the results obtained in the perturbation regime. For
comparison, the perturbation result $\Gamma _{1nl}^{\left( 2\right) }$ in
Eq. (\ref{8}) is depicted as the inset in Fig. \ref{fig2}(b), which is
exceeding the exact rate $\Delta \Gamma _{2}^{a}$. Therefore, $\Gamma
_{1nl}^{\left( 2\right) }$ serves as an upper bound for the
inhomogeneity-induced relaxation rate.

\begin{figure}[tbp]
\begin{center}
\includegraphics[width=0.85\linewidth]{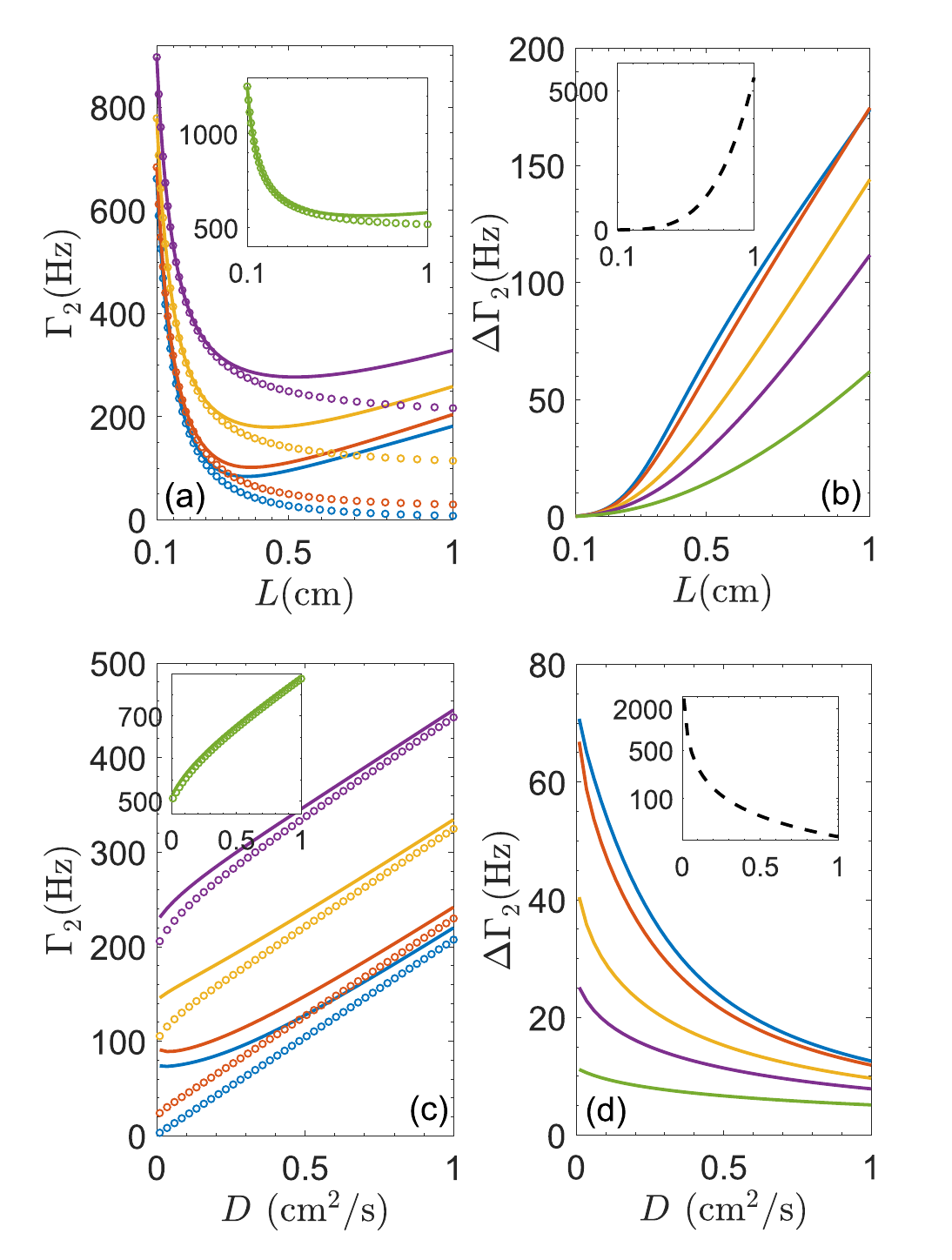}
\end{center}
\caption{(a) (c) Transverse relaxation rate $\Gamma _{2}^{a}$ and (b) (d)
the contribution $\Delta \Gamma _{2}^{a}$ from the field inhomogeneity. In
(a) and (c), solid curves represent $\Gamma _{2}^{a}$ with a linear gradient
of $\protect\gamma _{a}g=10^{3}$Hz/cm, while the circles are for $\protect%
\gamma _{a}g=0$. The diffusion constant $D_{a}$ for (a) and (b) is $0.2$cm$%
^{2}$/s and the length $L$ for (c) and (d) is $0.4$cm. The legend is the
same as in Fig. \protect\ref{fig2}, i.e., the decay rate $\Gamma _{0}^{a}\ $%
is taken to be $1$, $20$, $100$, $200$, $500$Hz. It increases from lower to
upper curves (solid lines and circles, respectively) in (a)(c) ($\Gamma
_{2}^{a}$ for $\Gamma _{0}^{a}=500$Hz are shown in the insets), while it
decreases from lower to upper curves in (b)(d). Also, the insets in (b) and
(d) give a upper bound of $\Delta \Gamma _{2}^{a}$ from the perturbation
theory.}
\label{fig3}
\end{figure}

We also investigate the variation of the transverse relaxation rate $\Gamma
_{2}^{a}$ with changing the system length $L$ and the diffusion constant $%
D_{a}$ vary, as depicted in Fig. \ref{fig3}, while maintaining a fixed
gradient $\gamma _{a}g=10^{3}$Hz/cm. In Fig. \ref{fig3}(a), $\Gamma _{2}^{a}$
initially decreases with $L$ since the effect from the wall collision \cite%
{Wu2021,PhysRevA.109.023113} surpasses that from the field inhomogeneity in
this range of $L$. As $L$ continues to increase, the field inhomogeneity
comes to dominate the relaxation, causing $\Gamma _{2}^{a}$ to increas with $%
L$. As a comparison, the ralaxation rate $\Gamma _{2}^{a}$ without the field
gradient is illustrated as circles in Fig. \ref{fig3}(a), representing a
monotonous function of the length $L$. Similar to Fig. \ref{fig2}, the
influence from the field inhomogeneity diminishes for larger $\Gamma _{0}^{a}
$, as shown in \ref{fig3}(b), where $\Delta \Gamma _{2}^{a}$ is upper
boundaed by $\Gamma _{1nl}^{\left( 2\right) }$. This trend persists in Figs. %
\ref{fig3}(c) and (d), where $\Gamma _{2}^{a}$ varies with $D_{a}$, showing
a decreases of $\Gamma _{2}^{a}$as $D_{a}$ increases due to motional
narrowing \cite{doi:10.1143/JPSJ.9.316,PhysRevLett.96.117203}.

\section{Line broadening for Serf magnetometers}

In this section, we examine the field-inhomogeneity-induced linewidth
broadening for SERF magnetometers \cite%
{Happer1977,Allred2002,PhysRevA.71.023405,Seltzer2008}, which have a narrow
linewidth, making their response to the inhomogeneous field $\vec{B}\left(
\vec{r}\right) $ more significant. In addition to $\vec{B}\left( \vec{r}%
\right) $, the magnetometer operates under a small transverse magnetic field
assumed to be $B_{y}\hat{e}_{y}$. Consequently,, the polarization $P_{x}$
along the $x$ direction, as a function of $B_{y}$, exhibits an approximately
Lorentzian lineshape $B_{y}/\left( B_{y}^{2}+w^{2}\right) $, where $w$ is
determined by half of the distance in $B_{y}$ between the minimal and
maximal values of $P_{x}\left( B_{y}\right) $. Given the absence of a large
magnetic field, both transverse and longitudinal relaxations are influenced
by the inhomogeneous field, contributing to the broadening of the linewidth $%
w$. In this context, we do not differentiate between relaxations in
different directions but concentrate on the overall effect on the linewidth $%
w$.

Similar to the last section, we assume that without the inhomogeneous field,
the longitudinal and transverse relaxation rates are the same, meaning that
the magnetometer works in the perfect Serf regime. Without loss of
generality, we consider a $\left. ^{87}\text{Rb}\right. $ magnetometer.
Consequently, the diffusive Bloch equation for the electron spin $\vec{S}%
\left( \vec{r}\right) $ in the long-term limit reads \cite%
{PhysRevA.58.1412,PhysRevA.109.023113}%
\begin{equation}
D_{a}\nabla ^{2}q\vec{S}\left( \vec{r}\right) +\gamma _{a}\left[ \vec{B}%
\left( \vec{r}\right) +B_{y}\hat{e}_{y}\right] \times \vec{S}\left( \vec{r}%
\right) -\Gamma _{0}^{a}\vec{S}\left( \vec{r}\right) +\frac{R}{2}\hat{e}%
_{z}=0,  \label{9}
\end{equation}%
where $\gamma _{a}$ is equal to the electron gyromagnetic ratio $\gamma _{e}$
and $q$ is the slow-down factor in the low-field limit \cite%
{Happer1977,Seltzer2008}. The field gradient is once again assumed to be
linear: $\vec{B}\left( \vec{r}\right) =\sum_{\alpha =x,y,z}\left( g_{\alpha
}\alpha -g_{\alpha }L/2\right) \hat{e}_{\alpha }$, where $g_{\alpha }$ is
the linear gradient in the direction $\alpha $ and an amount of $%
\sum_{\alpha =x,y,z}g_{\alpha }L/2$ is subtracted. This subtraction accounts
for the residual constant field induced by the linear gradient, and in
practical applications of SERF magnetometers, it is compensated by a uniform
magnetic field \cite{Seltzer2008,PhysRevA.109.023113}. The divengence of the
magnetic field $\nabla \cdot \vec{B}\left( \vec{r}\right) =0$, leading to $%
\sum_{\alpha =x,y,z}g_{\alpha }=0$. To analyze the contributions from the
gradient field in different directions, we first ignore this zero-divergence
requirement and consider the values $g_{\alpha }$ independently.

\begin{figure}[tbp]
\begin{center}
\includegraphics[width=0.85\linewidth]{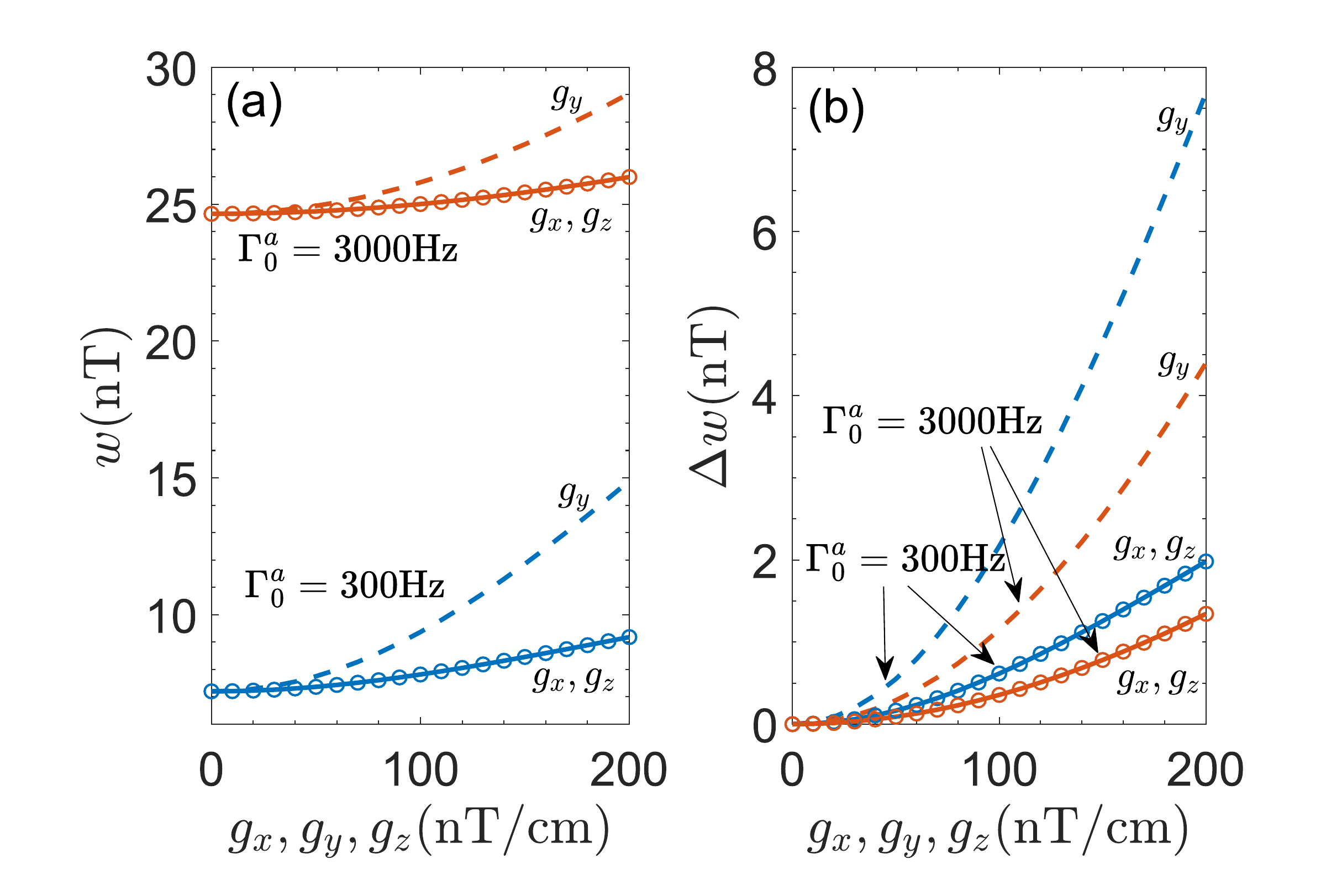}
\end{center}
\caption{(a) Linewidth $w$ and (b) linewidth broadening $\Delta w$ induced
by fied inhomogeneity with linear gradient. In (a), $\Gamma _{0}^{a}=300$Hz (%
$3000$Hz) for the two lower (upper) curves. The contribution to the
linedwith from the field gradient $g_{x}$ is the same as the one from $g_{z}$%
, while the linewidth broadening is much more sensitive to the inhomogeneous
field in the $y$ direction. }
\label{fig4}
\end{figure}

To acquire the linewidth, we solve Eq. (\ref{9}) numerically. The slow-down
factor $q$ is taken to be $6$, the value for $\left. ^{87}\text{Rb}\right. $
in the zero-polarization limit. Considering $D_{a}=0.2$cm$^{2}$/s and $L=0.2$%
cm, the dependence of the linewidth $w$ on the field gradient $g_{\alpha }$
for two values $\Gamma _{0}^{a}=300$Hz and $3000$Hz are shown in Fig. \ref%
{fig4}(a), along with the field-inhomogeneity-induced linewidth broadening $%
\Delta w\equiv w\left( g_{\alpha }\right) -w\left( g_{\alpha }=0\right) $ in
Fig. \ref{fig4}(b). Here, $w$ is determined by the linewidth of the average
spin polarization $\bar{S}_{x}\equiv \int dV$ $S_{x}(\vec{r})/L^{3}$.
Similar to the transverse relaxation induced by the longitudinal field
gradient, a smaller relaxation rate $\Gamma _{0}^{a}$ leads to larger
linewidth broadening. Here, the gradient in three directions are taken
account independently. In the plot, the curve marked with $g_{\alpha }$
indicates that the magnetic field $\vec{B}\left( \vec{r}\right) $ is $\left(
g_{\alpha }\alpha -g_{\alpha }L/2\right) \hat{e}_{\alpha }$ while its
components in other two directions are zero. The inhomogeneouse fields in
the $x$ and $z$ directions contribute equally to the linewidth (see the
proof in Appendix \ref{A1}), while the field along the $y$ direction have
much more significant influence on the linewidth broadening. The reason is
that the inhomogeneous field $B_{z}\left( \vec{r}\right) $ directly induces
dephasing in the $x$-direction spin $S_{x}\left( \vec{r}\right) $, and the
field $B_{x}\left( \vec{r}\right) $ induces dephasing in $S_{z}\left( \vec{r}%
\right) $ that is simultaneously transfered to the $x$-direction spins via $%
B_{y}$. However, the inhomogeneous field $B_{y}\left( \vec{r}\right) $ in
the $y$ direction directly affects both $S_{x}\left( \vec{r}\right) $ and $%
S_{z}\left( \vec{r}\right) $, resulting in a faster increase in the
linewidth.

\begin{figure}[tbp]
\begin{center}
\includegraphics[width=0.85\linewidth]{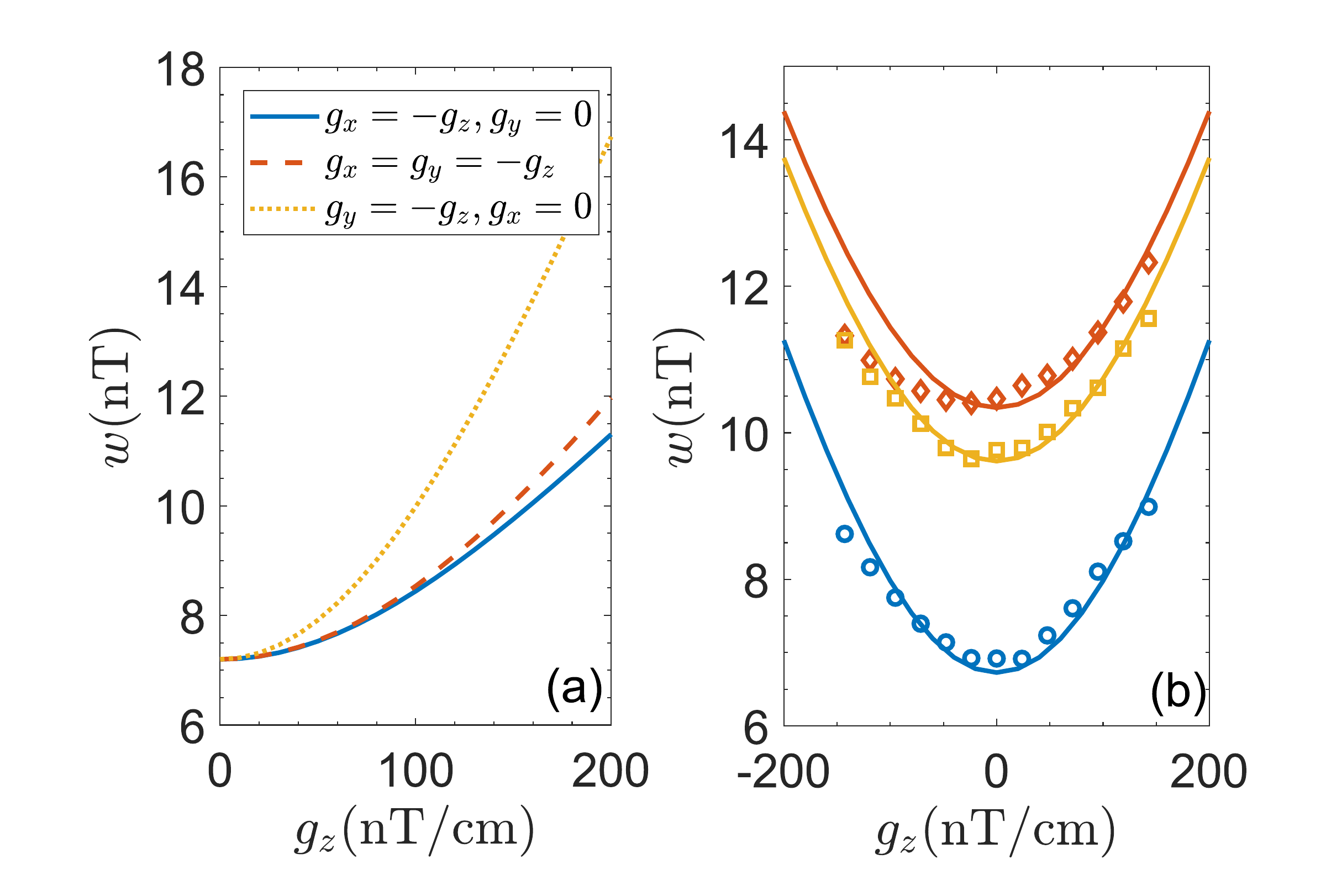}
\end{center}
\caption{(a) Linewidth $w$ with linear gradients $\sum_{\protect\alpha %
=x,y,z}g_{\protect\alpha }=0$ for three cases and (b) comparison between the
theoretical (shown in lines) and the experimental results (shown in circles,
diamonds and squares) under two different input powers. In (b), the pump
power for the circles (diamonds) is $50\protect\mu $W ($200\protect\mu $W),
and for the squares, the pump power is $200\protect\mu $W, but with a
detuning of $20$GHz in the pump laser. The solid lines are theoretical
results, with relaxation rates of $113$Hz, $562$Hz, and $675$Hz (from top to
bottom), corresponding to the circles, diamonds and squares, respectively.}
\label{fig5}
\end{figure}

Returning to the realistic case with $\sum_{\alpha =x,y,z}g_{\alpha }=0$, we
consider three representative conditions: (1) $g_{x}=-g_{z}$, $g_{y}=0$; (2)
$g_{x}=g_{y}=-g_{z}/2$; (3) $g_{y}=-g_{z}$, $g_{x}=0$. According to the
analyze above, the linewidth $w$ is largest in the third situation, as
denmonstrated in Fig. \ref{fig5}(a) with other parameters the same as in
Fig. \ref{fig4}. Therefore, in applications where the field gradient is
inevitable or difficult to compensate for, reducing the inhomogeneity in the
$y$-direction field can lower the dependence of the linewidth on the
gradient. We also compare our theoretical result with the experimental
observations using an uncoated $2$mm-cubic vapor cell. The cell, heated to $%
150^{\circ }$C, is filled with ankali-metal atoms $\left. ^{87}\text{Rb}%
\right. $ and $760$Torr nitrogen gas. The relaxation rate $\Gamma _{0}^{a}$
is varied by adjusting the input power of the pump laser from $50\mu $W to $%
200\mu $W. Accordingly, $\Gamma _{0}^{a}$ is set to $113$Hz and $675$Hz for
our theoretical calculations, with the slow-down factor $q=4.5$. The field
gradient in the three directions is taken as $g_{x}=g_{y}=-g_{z}/2$. As
shown in Fig. \ref{fig5}(b), the results align well with the experimental
data at low input power. However, at higher input power, a deviation
emerges, showing an asymmetric dependence of the relaxation rate on $g_{z}$
and $-g_{z}$. This deviation is likely caused by the light shift \cite%
{PhysRevA.58.1412,PhysRevA.99.063411}, which effectively behaves as an
inhomogeneous field along the $z$ direction and becomes significant at
higher input power. To verify it, we detune the frequency of the pump laser
by $20$GHz and acquire a more symmetric curve that better matched the
theoretical result. This suggests that, for large input power, the
propagation attenuation of the laser field needs to be considered.

\section{Conclusions and outlook}

In this study, we have investigated the relaxation of atomic spins in the
presence of inhomogeneous magnetic fields, particularly focusing on the
depolarizing-boundary condition. Unlike nuclear spins such as noble-gas
spins, which typically operate under a nondepolarizing condition, the
relaxation of atomic spins is influenced by the original relaxation rate $%
\Gamma _{0}^{a}$ in the absence of an inhomogeneous field. Notably, smaller $%
\Gamma _{0}^{a}$ leads to a more significant increase in the relaxation
rate, highlighting a dependence not observed in the nuclear spin scenario.

Utilizing perturbative calculations, we derived an analytical upper limit
for the transverse relaxation rate induced by a linear gradient field in the
longitudinal direction. Expanding our study to practical applications, we
connected these findings to the linewidth broadening observed in SERF
magnetometers---an application where field inhomogeneity plays a substantial
role. Our analysis revealed that the linewidth is most sensitive to the
inhomogeneous field along the magnetometer's sensitive axis.

While our theoretical results agree well with experimental data for small
input powers, a noticeable deviation was observed as the input power
increased in SERF magnetometers. This discrepancy suggests the need to
consider the Maxwell equations governing light propagation, particularly for
high input powers. In such cases, accounting for the nonuniformly
distributed light shift becomes crucial for a more accurate estimation of
the linewidth. This avenue opens up possibilities for further refinement and
enhancement of theoretical models when applied to real-world experimental
scenarios.

\acknowledgments This work was supported by the National Natural Science
Foundation of China under Grants No. U2141237 and No.62303068.

\appendix

\section{Equivalence of $B_{x}\left( \vec{r}\right) $ and $B_{z}\left( \vec{r%
}\right) $ on $\bar{S}_{x}$}

\label{A1}

In this section, we will prove that the $x$-direction average spin $\bar{S}%
_{x}$ is unchanged when exchanging $B_{x}\left( \vec{r}\right) $ and $%
B_{z}\left( \vec{r}\right) $. We start from Eq. (\ref{9}) and rewrite is as%
\begin{equation}
M\left(
\begin{array}{c}
S_{x}\left( \vec{r}\right) \\
S_{y}\left( \vec{r}\right) \\
S_{z}\left( \vec{r}\right)%
\end{array}%
\right) =\left(
\begin{array}{c}
0 \\
0 \\
\frac{R}{2}%
\end{array}%
\right) ,  \label{a1}
\end{equation}%
where the matrix%
\begin{equation}
M=\left(
\begin{array}{ccc}
D_{a}\nabla ^{2}q-\Gamma _{0}^{a} & -\gamma _{a}B_{z}\left( \vec{r}\right) &
\gamma _{a}B_{y}\left( \vec{r}\right) \\
\gamma _{a}B_{z}\left( \vec{r}\right) & D_{a}\nabla ^{2}q-\Gamma _{0}^{e} &
-\gamma _{a}B_{x}\left( \vec{r}\right) \\
-\gamma _{a}B_{y}\left( \vec{r}\right) & \gamma _{a}B_{x}\left( \vec{r}%
\right) & D_{a}\nabla ^{2}q-\Gamma _{0}^{a}%
\end{array}%
\right) .
\end{equation}%
Note that $M$ and $\vec{S}\left( \vec{r}\right) $ can be further written in
a basic, for instance, in the discrete coordinate space $\left\{
x_{i},y_{j},z_{k}\right\} $, or in the eigenmode $\psi _{mnl}^{\left(
0\right) }\left( x,y,z\right) $. Then, each elements of $M$ is actually a
matrix. In the Following, we take the basis $\psi _{mnl}^{\left( 0\right)
}\left( x,y,z\right) $ as an example, but the proof itself does not depend
on the choice of the basis.

In the basis $\psi _{mnl}^{\left( 0\right) }\left( x,y,z\right) $, the
diagonal operator $D_{a}\nabla ^{2}q-\Gamma _{0}^{a}$ is still diagonal,
since $\psi _{mnl}^{\left( 0\right) }\left( x,y,z\right) $ is the
eigenfunction of $\nabla ^{2}$ under the deporlarizing boundary condition.
We denote this diagonal matrix as $M_{0}$. Similarly, we denote a matrix $%
M_{\alpha =x,y,z}$ for $\gamma _{e}B_{\alpha }\left( \vec{r}\right) \ $and a
vector $v_{\alpha =x,y,z}$ for $S_{\alpha }\left( \vec{r}\right) $, where
the matrix elements $\left[ M_{\alpha }\right] _{mnl,m^{\prime }n^{\prime
}l^{\prime }}=\gamma _{e}\int dV$ $\psi _{m^{\prime }n^{\prime }l^{\prime
}}^{\left( 0\right) \ast }\left( x,y,z\right) B_{z}\left( \vec{r}\right)
\psi _{mnl}^{\left( 0\right) }\left( x,y,z\right) $ and $\left[ v_{\alpha }%
\right] _{mnl}=\int dV$ $\psi _{mnl}^{\left( 0\right) \ast }\left(
x,y,z\right) S_{\alpha }\left( \vec{r}\right) $. The average value $\bar{S}%
_{x}\equiv \int dV$ $S_{x}(\vec{r})/L^{3}$ can be written in the form of $%
v_{\alpha }$ as%
\begin{equation}
\bar{S}_{x}\equiv \frac{1}{L^{3}}v_{0}^{T}v_{x}
\end{equation}%
where $v_{0}$ is a vector with its element $\left[ v_{0}\right] _{mnl}=\int
dV$ $\psi _{mnl}^{\left( 0\right) \ast }\left( x,y,z\right) $. $v_{x}$ can
be acquired by solving the equation (\ref{a1}) and the average value%
\begin{equation}
\bar{S}_{x}=\frac{1}{L^{3}}\left(
\begin{array}{c}
v_{0} \\
0 \\
0%
\end{array}%
\right) ^{T}\left(
\begin{array}{ccc}
M_{0} & -M_{z} & M_{y} \\
M_{z} & M_{0} & -M_{x} \\
-M_{y} & M_{x} & M_{0}%
\end{array}%
\right) ^{-1}\left(
\begin{array}{c}
0 \\
0 \\
\frac{R}{2}v_{0}%
\end{array}%
\right) .  \label{a2}
\end{equation}%
With a unitary transformation%
\begin{equation}
U=\left(
\begin{array}{ccc}
0 & 0 & 1 \\
0 & 1 & 0 \\
1 & 0 & 0%
\end{array}%
\right) ,
\end{equation}%
$\bar{S}_{x}$ becomes%
\begin{eqnarray}
\bar{S}_{x} &=&\frac{R}{2L^{3}}\left(
\begin{array}{c}
v_{0} \\
0 \\
0%
\end{array}%
\right) ^{T}UU^{-1}\left(
\begin{array}{ccc}
M_{0} & -M_{z} & M_{y} \\
M_{z} & M_{0} & -M_{x} \\
-M_{y} & M_{x} & M_{0}%
\end{array}%
\right) ^{-1}\times  \notag \\
&&UU^{-1}\left(
\begin{array}{c}
0 \\
0 \\
v_{0}%
\end{array}%
\right)  \notag \\
&=&\frac{R}{2L^{3}}\left(
\begin{array}{c}
0 \\
0 \\
v_{0}%
\end{array}%
\right) ^{T}\left(
\begin{array}{ccc}
M_{0} & M_{x} & -M_{y} \\
-M_{x} & M_{0} & M_{z} \\
M_{y} & -M_{z} & M_{0}%
\end{array}%
\right) ^{-1}\left(
\begin{array}{c}
v_{0} \\
0 \\
0%
\end{array}%
\right)  \notag \\
&=&\frac{R}{2L^{3}}\left[ \left(
\begin{array}{c}
v_{0} \\
0 \\
0%
\end{array}%
\right) ^{T}\left(
\begin{array}{ccc}
M_{0} & -M_{x} & M_{y} \\
M_{x} & M_{0} & -M_{z} \\
-M_{y} & M_{z} & M_{0}%
\end{array}%
\right) ^{-1}\left(
\begin{array}{c}
0 \\
0 \\
v_{0}%
\end{array}%
\right) \right] ^{T}  \notag \\
&=&\frac{R}{2L^{3}}\left(
\begin{array}{c}
v_{0} \\
0 \\
0%
\end{array}%
\right) ^{T}\left(
\begin{array}{ccc}
M_{0} & -M_{x} & M_{y} \\
M_{x} & M_{0} & -M_{z} \\
-M_{y} & M_{z} & M_{0}%
\end{array}%
\right) ^{-1}\left(
\begin{array}{c}
0 \\
0 \\
v_{0}%
\end{array}%
\right) .  \label{a3}
\end{eqnarray}%
Comparing Eqs. (\ref{a2}) and (\ref{a3}), we can conclude that $\bar{S}_{x}$
is invariant when we exchange $M_{x}$ and $M_{z}$, proving the equivalence
of $B_{x}\left( \vec{r}\right) $ and $B_{z}\left( \vec{r}\right) $ on $\bar{S%
}_{x}$.

\bibliographystyle{unsrtnat}
\bibliography{ref}


\end{document}